# Explicit Gain Equations for Single Crystalline Photoconductors

Jiajing He[†ᵉ], Kaixiang Chen[†ᵉ], Chulin Huang[‡], Xiaoming Wang[†], Yongning He[‡] and Yaping Dan*[†‡]

†State Key Laboratory of Advanced Optical Communication Systems and Networks, University of Michigan – Shanghai Jiao Tong University Joint Institute, Shanghai Jiao Tong University, Shanghai 200240, China

‡School of Microelectronics, Xi'an Jiao Tong University, Xi'an, Shaanxi 710049, China

ᵉ These authors contribute equally.

*Email: yaping.dan@sjtu.edu.cn

**Abstract**

Photoconductors based on semiconducting thin films, nanowires and 2-dimensional (2D) atomic layers have been extensively investigated in the past decades. But there is no explicit photogain equation that allows for fitting and designing photoresponses of these devices. In this work, we managed to derive explicit photogain equations for silicon nanowire photoconductors based on experimental observations. These equations may be universal for photoconductors based on isotropic and covalently bonded low-dimensional semiconductors. The silicon nanowires were fabricated by patterning the device layer of silicon-on-insulator (SOI) wafers by standard lithography that were doped with boron at a concentration of ~ $8.6 \times 10^{17}$ cm$^{-3}$. It was found that the as-fabricated silicon nanowires have a surface depletion region ~ 32 nm wide. This depletion region protects charge carriers in the channel from surface scatterings, resulting in the independence of charge carrier mobilities on nanowire size. It is consistent with our Hall effect measurements but in contradiction with the accepted conclusion in the past decades that charge carrier mobilities become smaller for smaller nanowires due to surface scatterings. Under light illumination, the depletion region logarithmically narrows down and the nanowire channel widens accordingly. Photo Hall effect measurements show that the nanowire photoconductance is not contributed by the increase of carrier concentrations but the widening of the nanowire channel. As a result, a nanowire photoconductor can be modeled as a resistor in connection with floating Schottky junctions near the nanowire surfaces. Based on the photoresponses of a Schottky junction, we derived explicit photogain equations for nanowire photoconductors that are a function of light intensity and device physical parameters. The gain equations fit well with the experimental data, from which we extracted the minority carrier lifetimes as tens of nanoseconds, consistent with the minority carrier lifetime in nanowires reported in literature.

Photoconductors as the simplest photodetectors have been extensively investigated on various semiconducting materials in the past 60 years including thin films[1,2], nanowires[3-5], quantum dots[6,7] and more recently two-dimensional atomic layers[8-12]. The persistent research interests in photoconductors for decades are mainly driven by the extraordinarily high gain in quantum efficiency (up to $10^{10}$) observed in these photoconductors. However, there are still many fundamental issues unsolved in this field after decades of research. First, the classical gain mechanism predicts that high gain is intrinsic to all photoconductors.[2] It often misleads scientists to believe that high-gain and high-speed photoconductors can be made simply by shortening the device length[13], which however has never been realized. Recently we challenged this gain theory by pointing out that two assumptions in the derivation of the classical gain theory were misplaced.[14] The first assumption is that the classical theory assumes no metal-semiconductor boundary confinement, which leads to the questionable conclusion that high gain can be obtained as long as the minority recombination lifetime is much longer than the transit time. After the metal-semiconductor boundary confinement is considered, it turns out that a photoconductor intrinsically has no gain or at least no high gain.[14]

However, high gains in photoconductors are often observed in experiments.[3, 5, 15] The question is where these high gains are coming from.[16-18] This is because the classical theory incorrectly made a second assumption that the number of excess electrons and holes contributing to photoconductivity are equal. Although excess electrons and holes are generated in pairs, excess minority carriers are often trapped by defects or potential wells (depletion regions for instance) in semiconductors.[19,20] The same number of excess majority counterparts is accumulated in the conduction channel, leading to the experimentally observed high photogain.[14]

After the above two assumptions were corrected, we previously derived a gain equation in the same way as the classical gain equation was derived.[14] The gain equation can be used to conceptually explain experimental observations of photoconductors. However, similar to the questionable classical gain equation, the gain equation is not an explicit function of light intensity either but an expression of parameters that are difficult to directly calibrate and design. It is almost impossible to either design the photoconductive gain based on the gain theory or quantitatively check the gain theory against experimental data by fitting.

In this work, we managed to derive explicit photogain equations for single crystalline nanowire photoconductors in which the photogain in quantum efficiency is a function of light illumination intensity and other physical parameters including size, doping concentration and carrier mobility of underlying devices. These equations may be universal for photoconductors based on isotropic and covalently bonded low-dimensional semiconductors. The derivation is based on the experimental observations that the as-fabricated silicon nanowires have surface depletion regions. Light illumination creates a photovoltage across the depletion regions, which narrows down the depletion regions and therefore increases the channel width. Photo Hall effect measurements indicates that the nanowire photoconductivity is not contributed by the increase of carrier concentrations but the widening of the nanowire channel. As a result, a nanowire photoconductor can be modeled as a resistor in connection with floating Schottky junctions near the nanowire surfaces. The derived photogain equations fit well with the experimental data for all nanowires. From the fitting, we find the effective minority recombination lifetimes in the nanowires are approximately tens of nanoseconds, two orders of magnitude lower than that in bulk silicon with a *p*-type doping concentration of ~$10^{18}$ cm$^{-3}$. This is mainly due to the relatively high surface recombination velocity of silicon nanowires, and largely consistent with experimental observations reported in literature.[21]

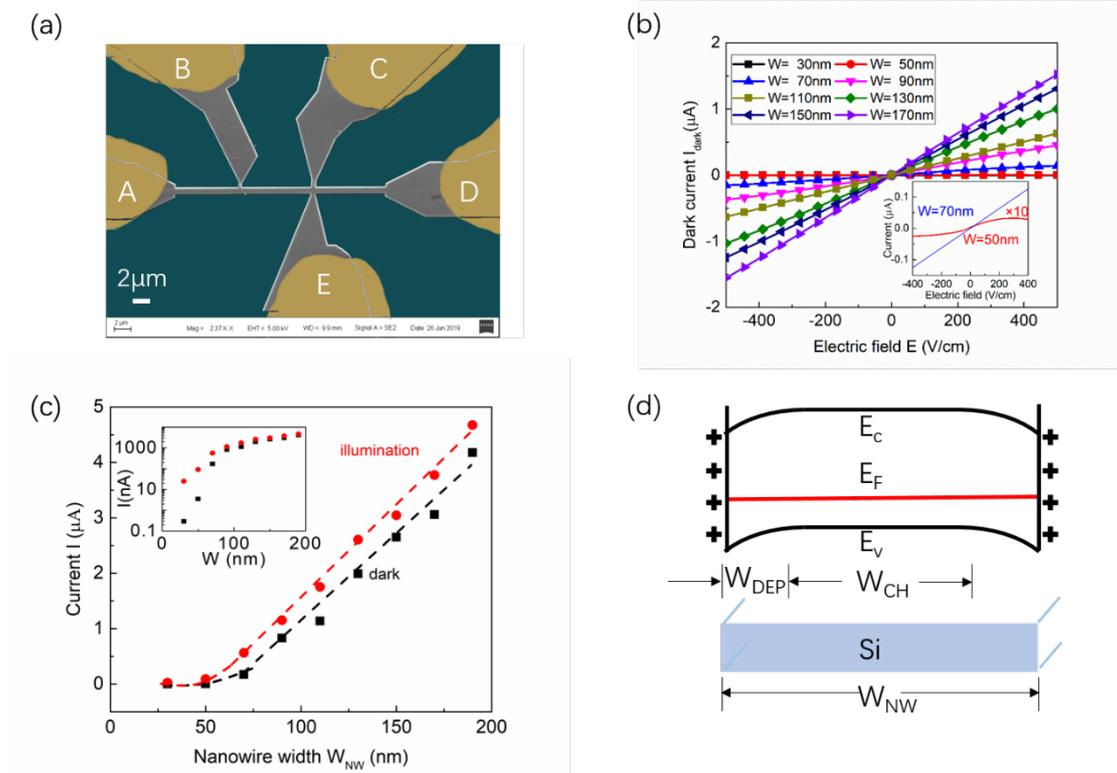

Figure 1. (a) False color SEM image of a silicon nanowire. Si, SiO$_2$ and Au are in gray, dark green and yellow, respectively. (b) I-V characteristics of typical silicon nanowires with a width ranging from 30 nm to 190 nm (four-probe measurements). (c) Current at a fixed bias for nanowires with different width in darkness and under light illumination (2mW/cm$^2$, $\lambda$=460nm). Inset: current in log scale. (d) Energy band diagram in the nanowire cross-section.

**Results and Discussion**

The silicon nanowires along with contact pads were fabricated by patterning the device layer (220nm thick) of a silicon-on-insulator (SOI) wafer that was heavily doped with boron at a concentration of ~$10^{18}$ cm$^{-3}$ by ion implantation (see Experimental Section for details). These nanowires are 27 μm long with a width ranging from 800 nm down to 30 nm. Figure1a shows the scanning electron microscopic (SEM) image of a silicon nanowire in contact with five pads for two-probe, four-probe and Hall Effect measurements. The two-probe and four-probe measurements show that the contacts of large pads are Ohmic and the contact resistances are negligibly small (See supplementary SI Section 1). The four-probe measured dark current vs electric field for nanowires with different width is shown in Fig. 1b. The current is linear with the applied electric field for the nanowires with a width larger than 60 nm, indicating that the nanowires have a uniform and continuous channel. For those narrower than 60 nm, the current is non-linearly dependent on the electric field (inset of Figure1b). The two-probe and four-probe measurements show that this nonlinearity comes from the nanowire channel instead of contacts.

Figure 1c exhibits the current at a fixed bias as a function of nanowire physical width in darkness (black squares) and under light illumination (red dots). In darkness, the current is linear with the nanowire width unless the width is smaller than $65 \pm 4$ nm, which is extracted from the extension of the linear dependence (Figure 1c). For those nanowires smaller than $65 \pm 4$ nm, the current follows an exponential dependence on the nanowire width as shown in the inset of Figure1c. The only possible explanation is that the nanowire channel is pinched-off by surface depletion regions as shown in the schematic of Figure 1d (see supplementary SI Section 2). The depletion region width is ~ 32 nm, half of the pinched-off nanowire width. Under light illumination, photogenerated electron-hole pairs will be physically separated by the built-in electric field in the surface depletion region, creating a forward photovoltage across the depletion regions. The forward photovoltage will consequently narrow down the depletion region width and widen the nanowire channel. This is consistent with our experimental observations in which the linear dependence of the nanowire conductance on the wire width is left-shifted under light illumination (red dots in Figure 1c).

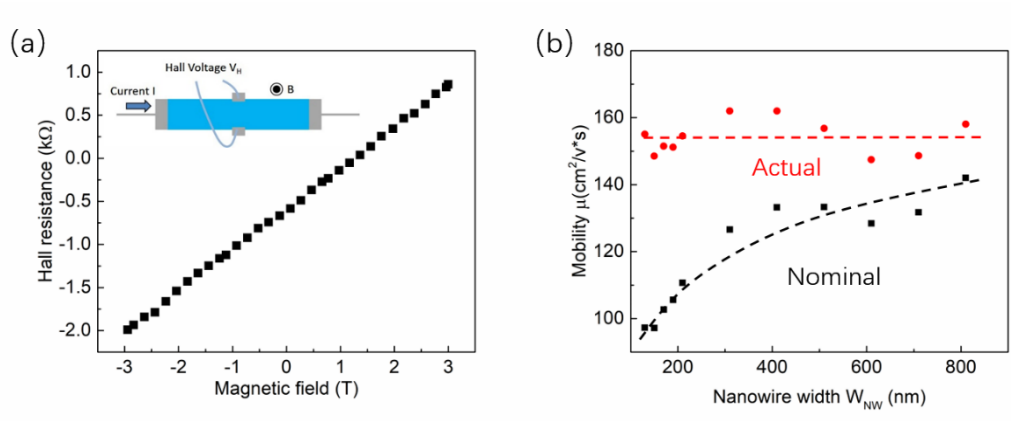

Figure 2. (a) Hall resistance vs applied magnetic field of photoconductors. Inset: Schematic of Hall effect measurements on nanowire device. (b) Hole mobility in nanowires with different widths when depletion regions are considered (red dots) or not considered (black squares).

The Hall effect measurements were conducted on the nanowires in a physical property measurement system (PPMS-9T, Quantum Design). The charge carriers transporting along the nanowire axial direction are being diverted towards the nanowire sidewalls by the magnetic field, generating a Hall voltage across the nanowire radial direction. The Hall resistance $R_H$ which is linear with the magnetic field can be written as eq.(1).

$$R_H = \frac{B}{qH_{ch}p} \quad \ldots (1)$$

,in which $B$ is the magnetic field, $q$ the unit charge, $H_{ch}$ the thickness of nanowire channel and $p$ the majority hole concentration in the channel. Considering that SOI wafers have a high-quality Si-SiO$_2$ interface, it is reasonable to assume that the depletion region exists only at the top surface and not at the bottom Si-SiO$_2$ interface. Given that the thickness of the device layer is ~ 200 nm (some thickness loss after surface cleaning) and the surface depletion region width was previously found as ~ 32 nm, logically $H_{ch}$ should be equal to ~ 168 nm. From eq.(1) and the linear correlation of Hall resistance and magnetic field, we found the doping concentration in the nanowires as ~ $8.6 \times 10^{17}$ cm$^{-3}$. It is known that the nanowire conductance σ is governed by eq.(2).

$$\sigma = qp\mu_p H_{ch}W_{ch}/L \quad \ldots (2)$$

, where $q$ is the unit charge, $\mu_p$ the majority hole mobility, $W_{ch}$ the channel width, H$_{ch}$

the channel height and L the nanowire length (9μm long between inner two electrodes in four-probe measurements). Since the conductance, hole concentration, physical dimension of the nanowires and surface depletion width (from Figure 1c) were measured, we can reliably find the hole mobility $\mu_p$ in the nanowire channel from eq.(2).

The results are plotted as the red dots in Figure 2b. It shows that the hole mobilities are independent of the nanowire widths, consistent with the fact that there are surface depletion regions near the nanowire surfaces, which keeps the nanowire channel away from the surfaces. The majority holes transport inside the nanowire channel and are free from the scattering of surface charges and roughness. However, it was widely reported that the carrier mobilities in semiconductor nanowires are dependent on the nanowire diameters.[22-24] This is because the surface depletion regions were not considered in these works. If we do not consider the surface depletion region either, the calculated hole mobilities will also decline as the nanowire becomes narrower (black squares in Figure 2b), similar to what was observed by others.[22-24] Given that charge carriers are free of surface effects, the dependence of carrier mobilities on doping concentration will follow the same correlation as in bulk semiconductors. The hole mobilities around 155 cm$^2$/Vs in *p*-type silicon corresponds to a doping concentration of ~1×10$^{18}$ cm$^{-3}$,[25] close to the doping concentration ~ 8.6×10$^{17}$ cm$^{-3}$ extracted from Hall effect measurements if the surface depletion regions in the nanowires are taken into consideration.

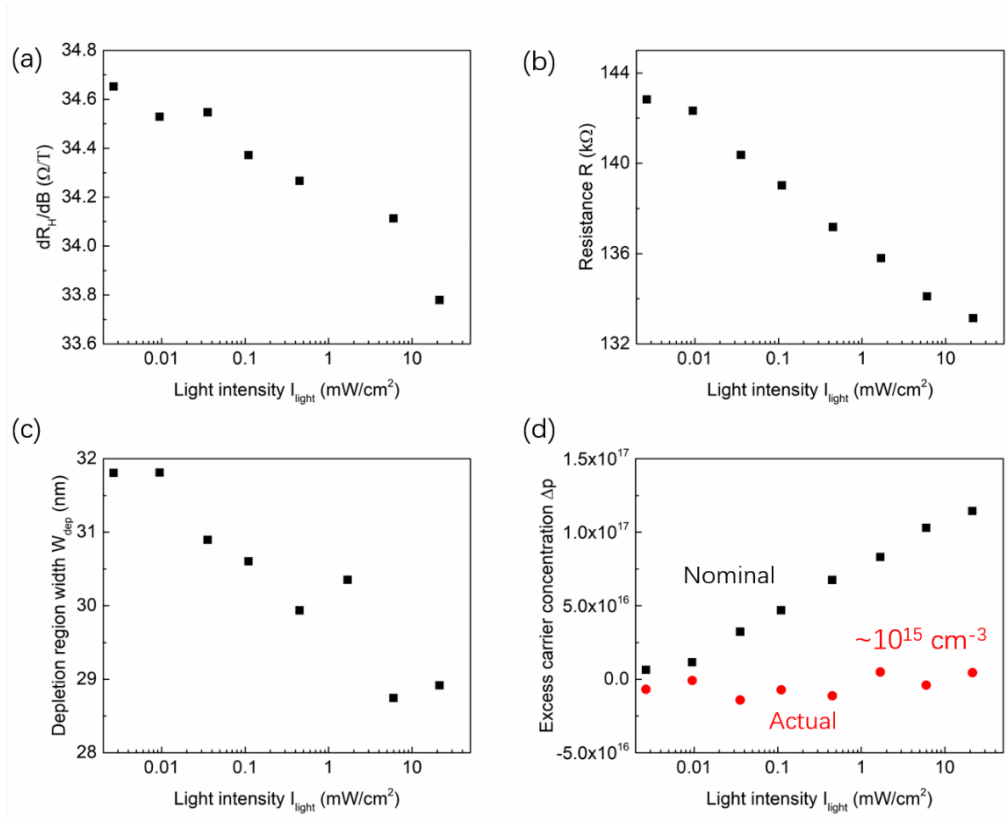

Figure 3 For a 160nm-wide nanowire device under the illumination of different light intensities (in vacuum at $\lambda$=532 nm), following parameters are measured and plotted as a function of light intensity: (a) derivative of Hall resistance respective to magnetic field, (b) nanowire resistance, (c) depletion region width, and (d) concentration of excess majority carriers when depletion region width is considered (red dots) or not considered (black squares).

Under light illumination, the photogenerated electrons and holes both contribute to photoconductivity. But our previous work[4] indicated that the excess majority concentration is orders of magnitude higher than the excess minority concentration in the channel ($\Delta p \gg \Delta n$ in our case). It is because excess minority electrons diffuse into the depletion region while excess majority holes are accumulated in the nanowire channel due to the built-in electric field of the surface depletion region. In such a case, eq.(1) and eq.(2) will still maintain its present form under light illumination except $p = p_0 + \Delta p$, $H_{ch} = H_{ch0} + \Delta H_{ch}$ and $W_{ch} = W_{ch0} + \Delta W_{ch}$ where "0" denotes the parameter in darkness. Under light illumination, the product of $H_{ch} \times p$ can be accurately found from eq.(1) using the derivative experimental data $\frac{dR_H}{dB}$ in Figure 3a.

After the product is plugged into eq.(2), the channel width $W_{ch}$ and then the depletion region width ($W_{dep}$) variation can be reliably found using the measured conductance under light illumination in Figure 3b. Figure 3c exhibits the depletion region width as a logarithm function of light intensity. In fact, from the product of $H_{ch} \times p$, we can find the excess hole concentration $\Delta p$ as shown in the black squares in Figure 3d on the assumption that $H$ does not change under light illumination. But light illumination will broaden both $H_{ch}$ and $W_{ch}$ at a similar magnitude due to the narrowing of the surface depletion region. When this narrowing is considered, the excess hole concentration $\Delta p$ will be at least two orders of magnitude lower (red dots in Figure 3d). This means that the light-induced phenomena in photoconductivity and photo Hall effect are dominantly contributed by the narrowing of the surface depletion region instead of the increase in excess majority carrier concentration, consistent with simulation results in literature. [26]

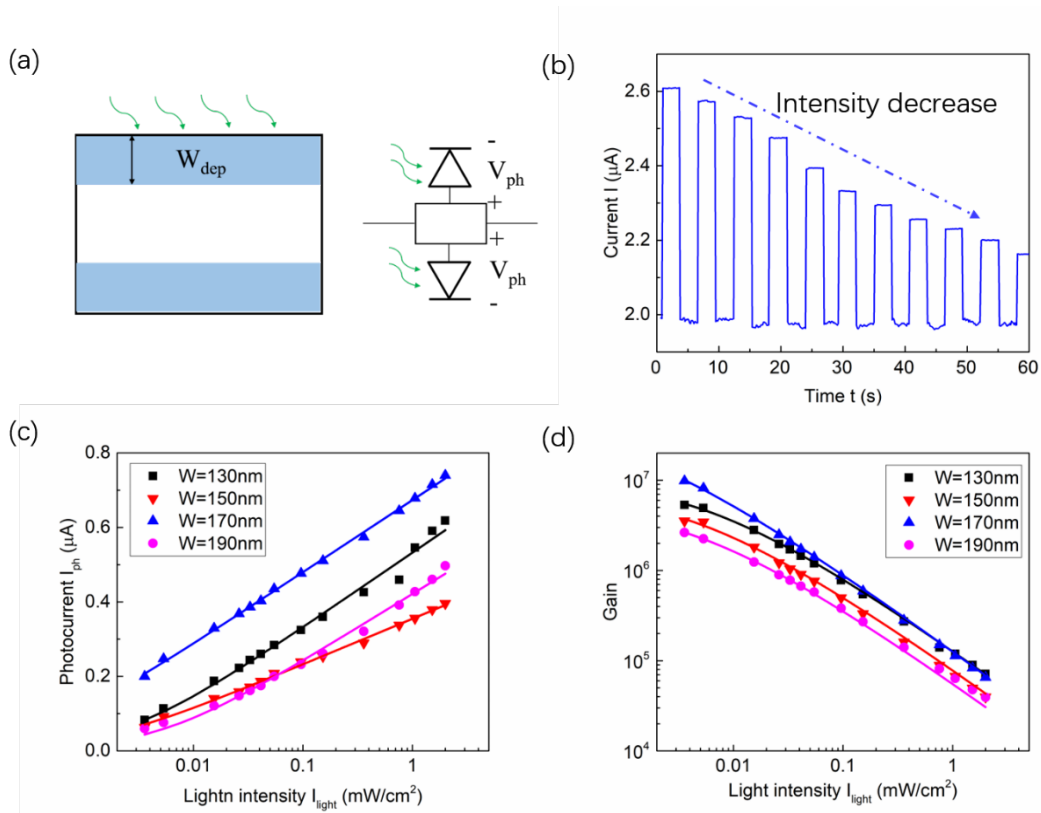

Figure 4. Device model and measured photoresponses of nanowire photoconductors in ambient at $\lambda$=460 nm. (a) Schematic of equivalent device model for photoconductors. (b) Current of a 130nm wide nanowire device under chopped illumination of different light intensities. (c) Measured photocurrent of nanowire devices as a function of light intensity. Solid lines are fitting curves following eq.(8). (d) Photogain of nanowire

photoconductors. Solid lines are fitting curves following eq.(9).

For this reason, a photoconductor with surface depletion regions can be modeled as a resistor in connection with floating Schottky junctions as shown in Figure 4a. The depletion region width of the Schottky junction controls the channel width of the resistor. Under light illumination, the photogenerated electron and hole pairs will be separated by the built-in electric field of the Schottky junction, creating a photocurrent $I_{ph}^c$ across the junction (perpendicular to the channel) as eq.(3).

$$I_{ph}^c = q \times g \times W_{dep} \times A_{junc} \quad \text{... (3)}$$

, where $q$ is the unit of charge, g the generation rate and $A_{junc}$ the cross-section area of the Schottky junction. The generation rate g is equal to the number of absorbed photons divided by the nanowire volume, i.e. $g = \frac{\alpha I_{light} A_{proj}}{\hbar \omega V_{NW}}$ in which $\alpha$ is the ratio of the photons absorbed by the nanowire to the total incident ones, $A_{proj}$ the area of the photoconductors projected in the light incident direction, $\hbar\omega$ the photon energy and $V_{NW}$ the nanowire volume. The absorption ratio α can be found by finite difference time domain (FDTD) simulations for each nanowire (see SI Section S3).

The photocurrent will be balanced by a forward current with a forward voltage (photovoltage) on the floating Schottky junction as shown in eq.(4).

$$I_{ph}^c = A_{junc} J_s [\exp\left(\frac{qV_{ph}}{nkT}\right) - 1] \quad \text{... (4)}$$

, where $J_s$ is the leakage current density of the Schottky junction, $q$ the unit charge, $k$ the Boltzmann constant, $T$ the absolute temperature and n the ideality factor. The light intensity can be derived as a function of the photovoltage by plugging eq.(3) into eq.(4):

$$I_{light} = I_{light}^S [\exp\left(\frac{qV_{ph}}{nkT}\right) - 1] \quad \text{...(5)}$$

, where $I_{light}^S = \frac{\hbar \omega J_s H}{\alpha W_{dep} q}$ is defined as the critical light intensity and $H$ is the nanowire physical thickness. The critical light intensity is the light intensity that creates a photocurrent density across the junction equal to the junction leakage current density.

It is known that the built-in potential ($V_{bi}$) depends on the depletion region width

($W_{dep}$) as $W_{dep} = \sqrt{\frac{2\epsilon_s V_{bi}}{qp}}$ in which $\varepsilon_{Si}$ is the silicon dielectric constant, $p$ the acceptor dopant concentration, and $q$ the unit of charge.[1] Previously we measured the depletion region width as ~ 32 nm and the doping concentration as ~8.6×10$^{17}$ cm$^{-3}$, from which the built-in potential ($V_{bi}$) can be calculated as 0.70 V. The forward photovoltage will narrow down the Schottky junction depletion region following eq.(6). If $\frac{\Delta W_{dep}}{W_{dep}} \leq 20\%$, the error of eq.(6) will be less than 10%.

$$\Delta V_{bi} = V_{ph} = \frac{2V_{bi0}}{W_{dep}} \Delta W_{dep} \quad \cdots (6)$$

The previous experimental data in Fig.3 show that the nanowire photoconductance is dominantly contributed by the nanowire channel widening. According to eq.(2), the nanowire photoconductance will be only a function of the channel cross-section area $A_{ch}$ (= $H_{ch} \times W_{ch}$ in our case) increase. Considering that no surface depletion exists at the bottom surface of the nanowire due to the high-quality Si/SiO$_2$ interface of SOI wafers, the derivative of channel cross-section area $A_{ch}$ respective to $W_{dep}$ is equal to $|\frac{dA_{ch}}{dW_{dep}}| = 2H_{ch} + W_{ch}$ in our case. For other device structures, this derivative can be found accordingly. The depletion region width variation can be written as

$$\Delta W_{dep} = \frac{I_{ph}}{\mu_p p q E_{ch} |\frac{dA_{ch}}{dW_{dep}}|} \quad \cdots (7)$$

, where $I_{ph}$ is the photocurrent, $E_{ch}$ the electric field intensity in the nanowire and other parameters have the same physical meaning as in eq.(2). By plugging eq.(7) into eq.(6) and then into eq.(5), we derive the photocurrent as a function of light intensity in eq.(8).

$$I_{light} = I_{light}^S \left[ \exp\left(\frac{2qV_{bi0}}{nkT} \frac{I_{ph}}{I_{ph}^S}\right) - 1 \right] \quad \cdots (8)$$

in which $I_{ph}^S = \mu_p p q E_{ch} W_{dep} |\frac{dA_{ch}}{dW_{dep}}|$ is defined as the critical photocurrent with the parameters given in previous equations. The critical photocurrent is the source-drain photocurrent when the depletion region $W_{dep}$ narrows down to zero by light illumination. Since all the parameters are known, the critical photocurrent can be calculated for different nanowires as shown in the second column of Table I.

Figure 4b exhibits the transient current responses as the light illumination intensity reduces. The resultant photocurrent dependence on light intensity is plotted in Figure 4c for four nanowires. Eq.(8) can fit well with the experimental data, from which we can extract the ideality factors n and the critical photocurrent $I_{light}^S$ as summarized in Table I. The critical light intensity $I_{light}^S = \frac{\hbar \omega J_s H}{\alpha W_{dep} q}$ is correlated with the absorption ratio α, the leakage current density of surface depletion region $J_s$ and the nanowire surface depletion region width $W_{dep}$. α can be found for all nanowires by performing FDTD simulations using commercial software Lumerical (5th column in Table I). At a given $W_{dep}$ (~32nm), we can calculate the leakage current density $J_s$ from the extracted $I_{light}^S$ for all four nanowires as listed in Table I. According to semiconductor device physics, the leakage current density $J_s$ in a depletion region can be further written as $J_s = \frac{q n_i W_{dep}}{2 \tau_0}$ in which $q$ is the unit charge, $n_i$ the electron concentration in intrinsic silicon, $W_{dep}$ the depletion region width and $\tau_0$ the effective minority carrier recombination lifetime in the doped silicon.[25] The effective minority carrier recombination lifetimes are found to be around tens of nanoseconds (except for the 170nm wide wire, last column in Table I), significantly smaller than the bulk minority carrier lifetime for Si at a doping concentration of $10^{18}$ cm$^{-3}$. This is caused by strong surface recombination in nanowires as reported in literature.[12, 27] The surface recombination velocity can further estimated on an order of $10^2$ cm/s from the nanowire dimensions and the effective minority carrier recombination lifetime.[12]

If the nanowire surface quality is repeatable and the doping concentration is uniform among nanowires, these fundamental parameters n and $\tau_0$ should have similar values from device to device. This is true except that n for the 150 nm wire is clearly smaller and $\tau_0$ for the 170 nm wire is much larger in comparison with the corresponding values of other devices.

For the 150 nm wire, n is smaller due to the clearly smaller slope of photocurrent vs light intensity (red triangles in Fig.4c). But it is unusual for *n* to be smaller than 1. The possible reason is that the calculated $I_{ph}^S$ is overestimated.

For the 170 nm nanowire, the minority carrier lifetime is much larger since the leakage current is small. When we examine the dark current for this 170 nm nanowire in Fig.1c, we find that the dark current is clearly below the dashed line, indicating that $W_{dep}$ for this nanowire is actually larger than the average ~ 32nm probably due to a higher concentration of surface states or charges. A careful analysis of Fig.1c shows that the 170nm nanowire has a depletion region width of ~37 nm, which gives a surface potential of ~0.9 V. The wider surface depletion region is consistent with the fact that the photocurrent for this nanowire is much larger than other nanowires (blue triangles in Figure 4c). The surface potential of ~0.9 V implies that the nanowire surfaces are in strong inversion and a high concentration of electrons is accumulated near nanowire surfaces. The inversion electrons will fill the surface states and reduce the surface recombination velocity (generation in depletion region). As a result, the leakage current density will be smaller.

**Table I** Calculated and Extracted parameters by fitting eq.(8) to the data in Fig.4c.

| NW width | $I_{ph}^S$ (μA) | n | $I_{light}^S$ (μW/cm²) | α | $J_s$(nA/cm²) | τ(ns) |
|---|---|---|---|---|---|---|
| 130 nm | 3.07 | 1.53 ± 0.06 | 2.28 ± 0.39 | 0.88 | 121.2 | 32.2 |
| 150 nm | 3.23 | 0.89 ± 0.02 | 1.29 ± 0.11 | 0.85 | 66.0 | 59.1 |
| 170 nm | 3.38 | 1.34 ± 0.02 | 0.33 ± 0.02 | 0.65 | 13.1 | 297.6 |
| 190 nm | 3.54 | 1.20 ± 0.04 | 4.79 ± 0.63 | 0.70 | 202.3 | 19.2 |

As a further step, we can find the photo gain $G$ from eq.(8).

$$G = \frac{I_{ph}/q}{I_{light}A_{proj}/\hbar\omega} = G_{max}\frac{I_{light}^S}{I_{light}}\ln\left(\frac{I_{light}}{I_{light}^S}+1\right) \quad \ldots(9)$$

with $G_{max} = \frac{n\hbar\omega kT I_{ph}^S}{2q^2 V_{bi0} A_{proj} I_{light}^S} = \frac{2\alpha\varepsilon_s nkT \left|\frac{dA_{ch}}{dW_{dep}}\right|}{q^2 n_i W_{dep} A_c}\frac{\tau_0}{\tau_t}$

, where $\alpha$ is the absorption ratio, $\varepsilon_s$ the dielectric constant of the underlying semiconductor, $n$ the ideality factor, $k$ the Boltzmann constant, $T$ the absolute temperature, $A_{ch}$ the channel cross-section area, $q$ the unit of charge, $n_i$ the intrinsic electron concentration of underlying semiconductor, $W_{dep}$ the surface depletion region width, $A_c$ the device physical cross-section area, $\tau_0$ the minority carrier recombination lifetime and $\tau_t$ the majority carrier transit time across the photoconductor which is equal to $\tau_t = \frac{L}{\mu_p E_{ch}}$ with L being the nanowire length, $\mu_p$ the majority carrier mobility and $E_{ch}$ the electric field intensity along the channel. Note that eq.(8) and (9) are valid on the condition that the electric field is not too strong to severely skew the excess carrier distribution due to the metal-semiconductor boundary confinement. Otherwise, the photocurrent will not be linear with applied electric field.[14]

The gain equation fits well with the experimental data shown in Figure 4d. As the light intensity $I_{light}$ gets weaker, the gain approaches to its maximum value $G_{max}$ according to eq.(9). Although the maximum photogain is also proportional to the ratio of the minority carrier recombination lifetime to the majority carrier transit time, the high gain does not come from the lifetime ratio but the channel width modulation by light illumination. This can be seen clearly if we take a closer look at the 130 nm nanowire for example. The maximum gain for this nanowire can be simplified as $G_{max} = 1.45 \times 10^6 \frac{\tau_0}{\tau_t}$ by plugging the related physical parameters into eq.(9). The minority carrier recombination lifetime is ~32.2ns and the carrier transit time ~5.3 ns. According to the classical gain theory[25], the excess carriers will circulate in the circuit ~ 6 times (=32.2/5.3) and the gain will be ~24 if the mobility difference is taken into account. But our gain equation predicts the photogain can be as high as ~$10^7$ if the light illumination is weak enough, consistent with our experimental measurements. Therefore, it is not surprising that Matsuo *et al.*[1] observed in 1984 that the gain of GaAs photoconductive detectors predicted by the classical gain theory is at least 3 to 4 orders of magnitude smaller than the gain measured in the experiments.

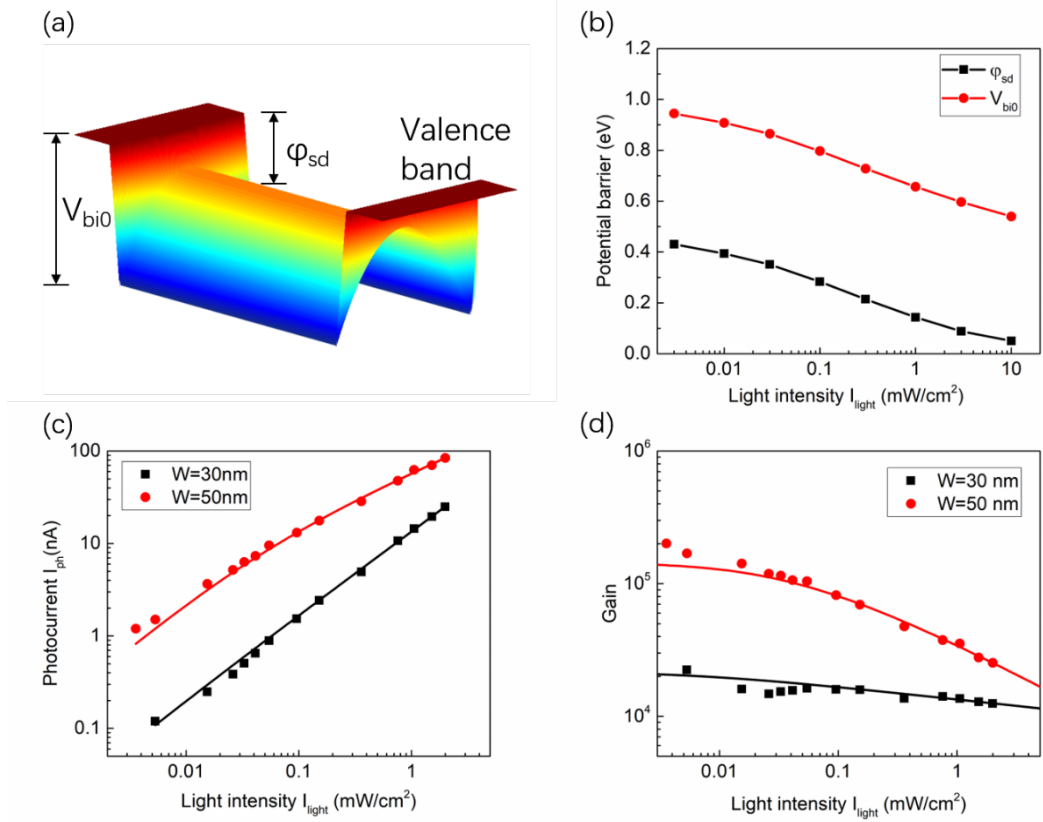

Figure 5. (a) 3D valence band profile of nanowires with a pinched-off channel. (b) Simulated potential barrier $V_{bi}$ and $\varphi_{sd}$ under illumination (see supplementary SI Section 3 for details). (c) Photocurrent current measured at $\lambda$=460 nm in ambient. Solid lines are fitting curves following eq.(13') (d) Photogain as a function of light illumination intensity. Solid lines are fitting curves following eq.(14).

For those nanowires narrower than 65 nm, the nanowire channel is pinched off by surface depletion region, creating a potential barrier $\varphi_{sd}$ between source and drain, like a metal-oxide-semiconductor field effect transistors (MOSFET) operating in subthreshold mode. Similar to the subthreshold current of MOSFET transistors, the current of our pinched-off nanowires also follows a similar correlation with the potential barrier as eq.(10).[25]

$$I \sim exp\left(-\frac{q\varphi_{sd}}{mkT}\right)\left[1 - exp\left(-\frac{qV_{sd}}{kT}\right)\right] \quad \ldots (10)$$

, in which $k$ is the Boltzmann constant, $T$ the absolute temperature, $q$ the unit charge and $V_{sd}$ the voltage bias between source and drain. $\varphi_{sd}$ is the minimal potential barrier between source and drain and m is a factor associated with the energy band bending from nanowire center to surface. Figure 5a exhibits the 3D energy band diagram for the valence band of the silicon nanowires. The potential barrier $\varphi_{sd}$ between source and drain is governed by eq.(11).

$$\varphi_{sd} = V_{bi0} - \frac{W_{dep}^2 qN_a}{2\epsilon_s} = V_{bi0} - \frac{W_{NW}^2 qN_a}{8\epsilon_s} \quad \ldots (11)$$

, in which $V_{bi0}$ is the potential between source/drain and the nanowire surfaces. Since the nanowires are completely pinched off, the depletion region width is equal to half the nanowire physical width, i.e. $W_{dep} = \frac{1}{2}W_{NW}$. The last term in eq.(11) is the potential difference from the nanowire center to surfaces. Clearly, the nanowire narrowing will increase the source-drain potential barrier $\varphi_{sd}$ and exponentially reduce the dark current, as show in Fig.1c. Interestingly, the light illumination will not narrow down the center-surface depletion region since it is limited by the nanowire physical width. As a result, $\varphi_{sd}$ is locked with $V_{bi0}$, resulting in the same magnitude of variation by light illumination (not too strong to open the channel) as shown in Figure 5b, *i.e.* $\Delta\varphi_{sd} \approx \Delta V_{bi0}$ (see supplementary SI Section 3 for details). This $\Delta V_{bi0}$ is the same $\Delta V_{bi0}$ in eq.(5) for nanowires with a continuous channel where the nanowire center-surface potential is equal to the source-surface potential ($V_{bi0}$). Therefore, we have eq.(12) by plugging $\Delta\varphi_{sd} \approx \Delta V_{bi0}$ into eq.(5).

$$I_{light} = I_{light}^S [\exp\left(\frac{q\Delta\varphi_{sd}}{nkT}\right) - 1] \quad \ldots (12)$$

, where n is the ideality factor which should be chosen ~1.3 according to Table I. From eq.(10) we can write $\frac{I_{ph}+I_{dark}}{I_{dark}} = \exp\left(\frac{q\Delta\varphi_{sd}}{mkT}\right)$. In the end, the photocurrent of photoconductors with a cut-off channel should be correlated with the incident light intensity as eq.(13).

$$I_{light} = I_{light}^S [\left(\frac{I_{ph}}{I_{dark}} + 1\right)^{m/n} - 1] \quad \ldots(13)$$

, which can be rewritten as $I_{ph} = I_{dark}[\left(\frac{I_{light}}{I_{light}^S} + 1\right)^{n/m} - 1] \quad \ldots(13')$

The resultant photocurrent as a function of light intensity is plotted in Figure 5c for two pinched-off nanowires. Eq.(13') can fit well with the experimental data. From the fitting, we extracted the critical light intensity $I^s_{light}$ and the factor n/m in Table II. Eq.(13) is valid on condition that the photocurrent is not too high to turn on the device, which is satisfied in our case. Similar to previous case, the leakage current J$_s$ and minority carrier lifetime can be found from $I^s_{light}$, the calculated α and the depletion region width W$_{dep}$ (~32 nm). The minority carrier lifetime is around tens of nanoseconds, comparable to what was found previously.

Eq.(13) is a power function and it is reasonable to plot it in a log-log scale. For very small nanowires, $I_{dark}$ is extremely small and $I_{ph}$ is often significantly larger than $I_{dark}$. Eq.(13) can be simplified as $I_{light} = I^s_{light}\left(\frac{I_{ph}}{I_{dark}}\right)^{m/n}$. The factor n/m is the slope of photocurrent vs light intensity in a log-log plot, which is often in the range of 0 and 1 as reported for nanowires and other nanomaterials in literature (m/n larger than 1 is also possible). For relatively large nanowires, $I_{dark}$ is not that small and the photocurrent can be much larger or smaller than $I_{dark}$ depending on light illumination intensity. The 50 nm wide nanowire is in this case. At the high light intensities, the dependence of photocurrent on light intensity is similar to smaller nanowires with the factor n/m as the slope. At low light intensities, $I_{ph}$ is much smaller than $I_{dark}$, as a result of which eq.(13) can be simplified as $I_{light} \approx I^s_{light}\frac{m}{n}\frac{I_{ph}}{I_{dark}}$. The slope of photocurrent vs light intensity in a log-log plot is approximately equal to 1.

**Table II** Extracted parameters by fitting eq(13') to Figure 5c.

| NW width | $I^s_{light}$(μW/cm²) | n/m | α | J$_s$(nA/cm²) | τ(ns) |
|---|---|---|---|---|---|
| 30nm | 2.62±0.21 | 0.90 | 0.60 | 94.33 | 41.1 |
| 50nm | 16.37±1.98 | 0.52 | 0.49 | 477.06 | 8.1 |

From eq. (13'), we can write the photo gain as eq. (14) which can fit well to the experimental data as shown in Figure 5d. As the light intensity reduces, the gain approaches a maximum value ($G_{max} = \frac{\hbar\omega}{qA_{proj}}\frac{n}{m}\frac{I_{dark}}{I^s_{light}}$) when the illumination light intensity is much smaller than the critical light intensity ($I_{light} \ll I_{light}^S$).

$$G = \frac{\hbar\omega}{qA_{proj}} \frac{I_{dark}}{I_{light}} \left[\left(\frac{I_{light}}{I_{light}^S} + 1\right)^{n/m} - 1\right] = G_{max} \left\{\frac{mI_{light}^S}{nI_{light}}\left[\left(1 + \frac{I_{light}}{I_{light}^S}\right)^{\frac{n}{m}} - 1\right]\right\} \quad \ldots (14)$$

, in which $G_{max} = \frac{n\hbar\omega I_{dark}}{mqA_{proj}I_{light}^S} = \frac{2n\alpha\tau_0 I_{dark}}{mqn_i V_{NW}}$ with $\alpha$ being the photon absorption ratio, $\tau_0$ the minority carrier recombination lifetime, $I_{dark}$ the dark current of the device, q the unit charge, $n_i$ the intrinsic electronic concentration of the underlying semiconductor and $V_{NW}$ the device physical volume. The maximum photogain $G_{max}$ is proportional to the minority carrier lifetime but irrelative to the carrier drift time. For the 50 nm wide nanowire, the maximum gain $G_{max}$ will be $1.6 \times 10^5$ although the minority carrier lifetime is as short as 8.1ns.

**Conclusions**

In this work, we successfully derived two explicit photogain equations for nanowire photoconductors with a continuous and pinched-off channel, respectively. These equations fit well with the experimental data and can be used to quantitatively explain and design photoresponses of nanowire photoconductors. These explicit gain equations clearly show that the high gain in photoconductors comes from the channel width modulation instead of excess carrier circulation. In terms of universality, these explicit gain equations may be applicable to isotropic and covalently bonded low-dimensional semiconductors in which interface defects and depletion regions play a dominant role in photoresponses. But the equations are unlikely applicable to 2D materials or other layered and van der Waals bonded structures due to their anisotropic nature.

**Experimental**

1. **Si nanowire fabrication**. The silicon-on-insulator (SOI) wafers were first cleaned with acetone and deionized (DI) water. Boron ions were then implanted into the device layer of the SOI wafer at an implantation energy of 30 keV and a dose of 2.2×10$^{13}$ cm$^{-2}$. The doping peak is located at 110nm with a maximum concentration of ~1×10$^{18}$ cm$^{-3}$. After the implantation, rapid thermal annealing (RTA) were employed to activate the boron atoms at 1000 °C for 20s. Afterwards, the silicon wafers were cut into small pieces (~ 1 × 1 cm$^2$). Each piece was cleaned with piranha solutions (98% $H_2SO_4$:30% $H_2O_2$, 3:1 (v/v)). A 250nm-thick layer of Polymethyl methacrylate (PMMA) resist (XR-1541-006, Dow Corning Electronics, USA) was spin-coated on the *p*-type SOI samples at 4000rpm for 60 s. Then, the wafers were baked at 180 °C for 90 s. The PMMA resist was exposed by electron beam lithography (Vistec EPBG5200) and subsequently developed in MIBK and IPA. PMMA is a positive electron beam resist. After that, aluminum (Al) was evaporated to the exposed region. To form etch mask for the micropads, NR9-1500PY (Futurrex Inc. USA) photoresist was coated on the wafers at 4000 rpm for 40 s. After baked at 140 °C for 60 s, the NR9 resist was exposed to UV light (MDA-400) and developed in the developer after post baking at 110 °C for 60 s. 100nm thick Al film was evaporated (Denton multi-target Magnetic Control Sputtering System) followed by liftoff process. The Al nanowire and micropad patterns defined by electron beam exposure and photolithography were then transferred to the SOI device layer by reactive ion etching (RIE, Sentech ICP Reactive Ion Etching System), forming an array of silicon nanowires with each connecting to two silicon pads for metal contacts. After photolithography, 20 nm thick Cobalt and 180 nm thick Aluminum were evaporated onto the Si pads as electrodes. To form good Ohmic contacts, the wafers were then annealed in Argon atmosphere for 15 min at 230 °C.

2. **Optoelectronic measurements**. The devices were characterized in a probe station by high-precision digital sourcemeters (Keithley 6487 and 2636). Light from a commercial LED ($\lambda$=460nm) forms a uniform light illumination spot (~5mm in diameter) on the samples. The light intensity is controlled by the driving current of the circuit. The light intensity is calibrated by a commercial photodiode (G10899-003K, Hamamatsu).

**Supporting information**

   **Four-probe and two-probe measurements; Simulated potential profile for nanowires with a continuous channel; FDTD simulation; Simulated potential profile for nanowires with a cut-off channel.**

**Acknowledgement**

The work is financially supported by the special-key project of the "Innovative Research Plan", Shanghai Municipality Bureau of Education (2019-01-07-00-02-E00075) and National Science Foundation of China (NSFC) (No. 61874072). The devices were fabricated at the center for Advanced Electronic Materials and Devices (AEMD) and Hall measurements were conducted at the Instrumental Analysis Center (IAC), Shanghai Jiao Tong University.

**Author Contributions**

Y. D. conceived the idea and derived the theoretical equations. J. H., K. C. and Y. D. analyzed the data. Y. D., J. H. and K. C. wrote the manuscript. J. H., K. C., C. H. conducted the experiments. C. H. and X. W. performed electronic and optical simulations. Y. H. commented on the manuscript. All authors reviewed the manuscript.